\documentclass[12pt]{iopart}
\usepackage{graphicx}

\usepackage[colorlinks,linkcolor=blue,anchorcolor=blue,citecolor=blue,urlcolor=blue,dvipdfm]{hyperref}
\usepackage{graphics}
\usepackage{stmaryrd}
\usepackage{float}
\usepackage{array, color}
\usepackage{tabularx}
\usepackage{threeparttable}
\usepackage{titletoc}
\usepackage{booktabs}

\usepackage{subfigure}

\usepackage{latexsym,bm}

\begin{document}

\title[Structure, equation of state, diffusion and viscosity of warm dense Fe]{Structure, equation of state, diffusion and viscosity of warm dense Fe under the conditions of giant planet core}

\author{Jiayu Dai$^{1,3}$, Yong Hou$^1$, Dongdong Kang$^1$, Huayang Sun$^1$, Jianhua Wu$^1$, Jianmin Yuan$^{1,2,3}$}

\address{$^1$Department of Physics, College of Science, National University of Defense Technology, Changsha 410073, P. R. China}
\address{$^2$State key laboratory of high performance computing, National University of Defense Technology, Changsha 410073, P. R. China}
\address{$^3$Author to whom any correspondence should be addressed.}
\ead{jydai@nudt.edu.cn} \ead{jmyuan@nudt.edu.cn}
\date{\today}

\begin{abstract}
Fe exists abundantly in the universe. In particular, the dynamical structures and transport properties of warm dense Fe are crucial for understanding the evolution and structures of giant planets. In this article, we present the ionic structures, equation of states, diffusion and viscosity of Fe at two typical densities of 33.385 g/cm$^3$ and 45 g/cm$^3$ in the temperature range of 1 eV and 10 eV, giving the data by the first principles calculations using quantum Langevin molecular dynamics (QLMD). Furthermore, the validation of Stokes-Einstein (SE) relation in this regime is discussed, showing the importance of choosing the effective atomic diameter. The results remind us of the careful usage of the SE relation under extreme conditions.
\end{abstract}

\tableofcontents
\maketitle

\section{Introduction}

The properties of complex materials such as Fe under extreme conditions are crucial for understanding the evolution of planets (earth, giant planets, extro-solar giant planets) \cite{apj1} and stars such as sun \cite{apj2,gao}. More interestingly, the laser-driven and Z-pinch experiments can lead to very high densities up to hundreds of g/cm$^3$ and temperatures up to thousands of eV \cite{nif,laser}. In particular, the temperatures of 1-10 eV with associated pressures of 10-1000 TPa are likely to exist in the interiors of massive planets \cite{fe1,fe2}. Recent studies have shown the possible stable structures at these pressures \cite{fe1,fe2}, where face-centered cubic (fcc) structures are stable in the range of 7-21 TPa. However, the effect of temperatures on these structures is still not known, since the dynamical structures such as melting, diffusion and viscosity can be induced by temperatures, and the phases of Fe in earth-like exoplanets are likely to be liquid. In order to understand these behaviors, the experimental determination seems currently extremely expensive and difficult. Therefore, accurate simulations are required for the determinations of the structures and dynamics. The dynamical structures of Fe and its compounds from first-principles molecular dynamics (FPMD) at the physical conditions of the Earth's core have been studied widely \cite{tran1,tran2,tran3}, showing the complexity of the new phenomena in high energy density physics (HEDP). In particular, it is worthy to verify whether the Stokes-Einstein (SE) relation at so high densities is valid, since SE relation is one important bridge between dynamics and statistics, and usually adopted in the previous calculations of the viscosities from diffusion coefficients \cite{tran1,tran2,tran3,tran4,tran5}. Up to now, the diffusion and viscosity can be obtained directly by molecular dynamics simulations with Kubo-Green relation \cite{tran3,tran4,tran5,Saigo}. Moreover, we can construct statistical models according to the simulation results, introducing easy applications in hydrodynamics. \cite{murrilo,Daligault}, However, molecular dynamics simulations are strongly dependent on the accuracy of many-body interatomic potentials, and there are usually some limitations for the statistical methods.

Cold stable fcc structure with its inducing liquid at high temperature is considered as an important phase in the giant planets such as Jupiter and exoplanets \cite{fe1,fe2}. In this regime, the densities are from 33.9 g/cm$^3$ to 66.7 g/cm$^3$, and the temperatures are from 1 eV to 10 eV. For these states, the melting behaviors and the transport properties are interesting and deserved to be studied carefully. Besides, much high density will introduce much different chemical bonds since the pressure larger than 1 Mbar will change the traditional chemistry dramatically \cite{dense1,dense2,dai1,dai2,dai4}. Furthermore, when the pressure increases to 100 Mbar, the core electron charge density can be changed significantly \cite{dense1,dense2,dai1}. Therefore, the transport properties such as equation of states (EOS), diffusion and viscosity will change a lot since these properties are related to the electronic structures. With respect to the methods of calculating these properties accurately, the first principles calculations are thus required because we did not understand their behaviors from semiclassical models such as average atom (AA) model \cite{aayuan,aamd1} and Thomas-Fermi (TF) model or orbital free (OF) method \cite{TF1,TF,TF2}, in which the orbital behaviors or chemical bonds information can not be described well.

Molecular dynamics combining finite-temperature density functional theory (DFT) \cite{Mermin} called quantum molecular dynamics (QMD) and quantum Langevin molecular dynamics (QLMD) have been successfully used in warm and hot dense matter, especially under very high pressures \cite{tran1,tran3,dai1,dai2,dai4,s2,dai5}. The validations of QMD and QLMD at high densities and temperatures have been verified a lot of times by comparing with experiments and interestingly reasonable results are found. In particular, QMD is considered as one of the most promising methods and has been successfully used to predict the transport properties of warm dense matter (WDM).\cite{wdm2,CH} The biggest argument for QMD applications is that its limitation on the number of particles because of its computational cost at high temperature. However, QMD and QLMD can be efficient for the calculations of WDM under the conditions of giant planet core, and give relatively accurate transport properties using a few hundred particles.\cite{wdm2,LiH,CH,Pu} In this article, we adopted QLMD to simulate the transport properties of dense Iron at 45 g/cm$^3$ from 1 eV to 10 eV. Except for, considering the density of 33.385 g/cm$^3$ at 100 eV along the Hugoniot curve \cite{dai1}, we also calculate the dynamical properties at 33.385 g/cm$^3$ up to 10 eV. Furthermore, the approximation of SE relation is discussed, showing the debatable validation under so dense regimes.

\section{Methods}

For the complexity of Fe at high density, both QMD with velocity rescaling method and QLMD simulations are tested firstly. In fact, at low temperature, the electron-ion collisions induced friction (EI-CIF) is small since the friction is proportional to the temperature. Therefore, QMD and QLMD are equivalent for the equilibrium properties at low temperature. Eventually, the advantage of QLMD is the higher computational efficiency than traditional QMD \cite{dai3}. The electronic structures are solved based on finite temperature DFT implemented in the Quantum ESPRESSO package \cite{qe}. Considering the small atomic sizes at high density, the time step of 0.5 fs is used in order to keep the correct trajectories of ions. Pseudopotential (PP) is one of the key points in QMD and QLMD simulations. Here, we construct a new PP with 16 electrons in the valence and 0.9 atomic units for the radius cutoff within the generalized-gradient approximation (GGA) \cite{gga}, which can promise the correctness with respect to the conditions we are studying. Using this PP, we can reproduce the bulk modulus properties of Fe as previous results calculated by full-electron calculations and experiments. More importantly \cite{bulk,dai1}, at high pressure, the pressure and band structures at the density of 33.9g/cm$^3$ and 48.23g/cm$^3$ are in good agreement with the pressure in Ref.~\cite{fe2} using the same PP \cite{dai1}. Furthermore, the same PP has been successfully used to calculate the EOS of Fe in hot dense regime within a wide range of densities and temperatures \cite{dai1}, indicating the validation of PP in this work. Besides, 2000 time steps with a large convergent tolerance of 1.0$\times 10^{-4}$ in electronic structure calculations are adopted in order to reach the thermalization, and 10 ps time lengths with a small convergent tolerance of 1.0$\times 10^{-6}$ \cite{dai3} are simulated to obtain the transport properties and thermal average. In order to accelerate the calculations, the Gamma point is only used for the representation of Brillouin zone. For comparing with the results from semiclassical methods, averaged atom molecular dynamics (AAMD) \cite{aamd1,aamd2} is performed for all cases. When two atoms are close enough, their electronic distributions would be overlapped, inducing interactions between them. In order to describe this interactions, we should establish the interatomic potential. In AAMD method, the temperature-dependent pair potential, which can describe the contributions of the overlapped electronic distributions, is constructed from the AA electronic calculations, including the ionic correlations within the framework of AA model. Within this framework, the ions are moving on this pair potential, giving the ionic trajectories and ionic configurations at different times. From AAMD, we can basically use large number of particles and giving the convergent results corresponding to the size effects. The comparison between QLMD and AAMD has been reported before \cite{dai2,dai3}, showing their agreement at high temperatures. The details of AAMD can be found in Ref.~\cite{aamd2}.

The self-diffusion coefficient D can be obtained from two methods. One is from the trajectories by the mean-square displacement $D^{R}$
\begin{equation} \label{e1}
D^{R}=\frac{1}{6t}<|\textbf{R}_i(t)-\textbf{R}_i(0)|^2>
\end{equation}
where the $\textbf{R}_i$ is the position of the $i_{th}$ particle. The other method is from the velocity autocorrelation function $D^{V}$
\begin{equation} \label{e2}
D^{V}=\frac{1}{3}\int_o^\infty <\textbf{V}_i(t)\cdot \textbf{V}_i(0)>dt
\end{equation}
where $\textbf{V}_i$ is the velocity of the $i_{th}$ particle. Besides, the viscosity $\eta$ can be obtained from the autocorrelation function of the off-diagonal component of the stress tensor so called Green-Kubo equation \cite{tran3,gk}
\begin{equation} \label{e3}
\eta=\lim_{t\rightarrow\infty}\overline{\eta}(t)=\lim_{t\rightarrow\infty}\frac{V}{k_BT}\int_o^t <P_{12}(0)\cdot P_{12}(t^{'})>dt^{'}
\end{equation}
where $P_{12}$ represents the averaged result for the five independent off-diagonal components of the stress tensor $P_{xy}$, $P_{yz}$, $P_{zx}$, ($P_{xx}-P_{yy}$) / 2, and ($P_{yy}-P_{zz}$) / 2. Using this method, the self-diffusion coefficient, viscosity of warm and hot dense matter have been reported widely \cite{tran1,tran2,tran3,tran4,tran5,wdm2,wdm1,wdm3,wdm5,wdm6} within the framework of QMD, showing its validation in these extreme conditions.

On the other side, the viscosity can also be calculated directly from the SE relation
\begin{equation} \label{e4}
D\eta=k_BT/(2\pi a)
\end{equation}
where $a$ is an effective atomic diameter. This relation is statistically obtained from the Brownian motion of a macroscopic particle in liquid, but it is only an approximation for the atoms. If the validation of SE relation can be verified, the size of particles, the transport behaviors can be understood well \cite{SE1}. In fact, for a Brownian particle, SE relation is equivalent to Eq.~\ref{e2} and Eq.~\ref{e3}. In particular, for the dense matter, the diameters of atoms are very small, and the validation of SE relation should be examined very carefully.

\section{Convergence tests}

\begin{figure}[!tb]
\centering
\includegraphics*[width=3.2in]{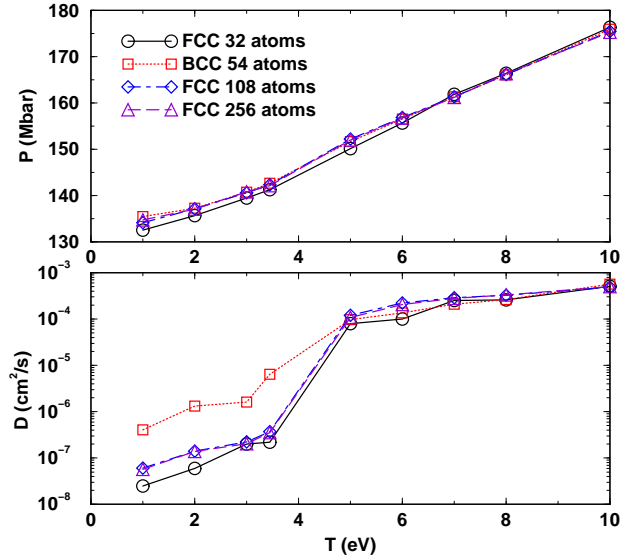}
\caption{(Color online) Size effect on the pressure (upper panel) and diffusion coefficient (lower panel) of Fe at 45 g/cm$^3$.} \label{fig1}
\end{figure}

For calculating the transport properties from first principles, the number of atoms should be tested since diffusion and viscosity are strongly dependent on the system sizes. For this purpose, in QLMD simulations, we tested face-centered cubic (FCC) structures with 32, 108 and 256 atoms and body-centered cubic (BCC) structure with 54 atoms in a supercell at 45 g/cm$^3$. The pressure and diffusion versus temperature relations with different number of atoms are shown in Fig.~\ref{fig1}. It is concluded that the pressure is not much sensitive to the system sizes. In particular, the pressures are closer for different sizes when the temperatures are higher. The small difference at low temperatures may be caused by the long ranged orders with crystal structures. It is worth noting that the initial configurations of BCC and FCC structures are important for the calculation of diffusion in the solid phases, but the differences from different initial configurations disappear gradually with the increasing temperatures in the liquid phases (here above 5 eV).

According to the tests, it can be known that our systems are really convergent for the calculation of pressure and diffusion coefficients when we use 108 atoms with initial FCC structures. It should be noted that when the system becomes liquid (after the jump in diffusion coefficient), the results of 54 atoms, 108 atoms and 256 atoms are consistent. The difference at low temperatures would be from the different initial structures of BCC or FCC crystals. In fact, the small differences from different sizes can also come from the statistical errors for small sizes. Basically, this error can be overcome from the long time simulations.

\begin{figure}[!b]
\centering
\includegraphics*[width=3.2in]{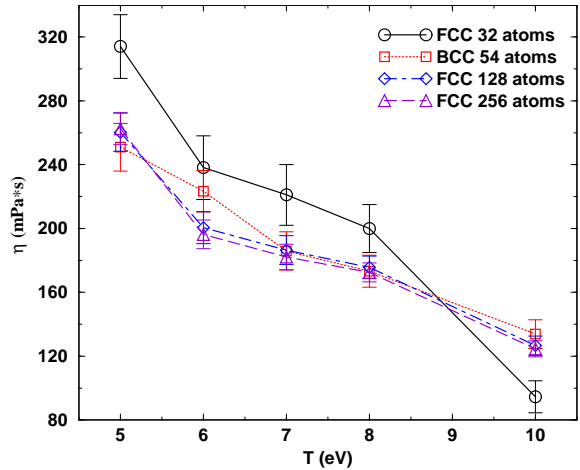}
\caption{(Color online) Size effect on the viscosity of Fe at 45 g/cm$^3$ from QLMD simulations.} \label{fig2}
\end{figure}

For the viscosity, the results of the convergent tests are shown in Fig.~\ref{fig2}. It is only useful for the liquids here since the viscosity is very large for solids. Here, the liquid phases are shown above 5 eV in our simulations. By comparison, we can know that the viscosity are much sensitive to the sizes. For small size of 32 atoms, the statistical errors are really large which is not likely to be compensated by increasing the simulation time. When the number of atoms is increased up to 108 and 256, the viscosities at different temperatures seem convergent. Therefore, we can safely use the 108 atoms for the calculations of pressure, diffusion, and viscosity in our cases. In fact, about one or two hundred atoms are usually used in QMD simulations for the transport properties\cite{tran1,tran3,tran5,wdm2,LiH,CH}, which has been shown to be convergent within reasonable errors.

\begin{figure}[!tb]
\centering
\includegraphics*[width=3.2in]{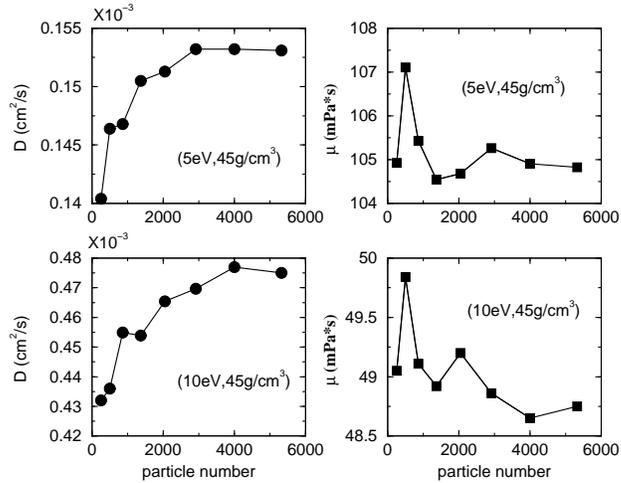}
\caption{(Color online) Size effect on the viscosity of Fe at 45 g/cm$^3$ from AAMD simulations.} \label{aare}
\end{figure}

For the effect of the particle number, we use large number of particles up to 5324 in AAMD calculations to verify the convergence, as shown in Fig.~\ref{aare}. The simulation time is up to 100 ps in each case in order to obtain the correct correlations of particles in these semiclassical calculations. It can be shown that although the number of particles can affect the convergence of diffusion and viscosity, all the differences are within 10\% or smaller. When the number is reach to 4000, the diffusion and viscosity are almost convergent. In order to compare the results, we use 4000 particles in AAMD to calculate the transport properties below. It is worth noting that AAMD adopted pair potentials, which should be more sensitive to the size and argued for the viscosity since the unsymmetrical shear of the system can not be described well by pair potentials.

For the density of 33.385 g/cm$^3$, we used 54 atoms with initial BCC structures to calculate their transport properties. The simulation details are also the same as those of 45 g/cm$^3$. Basically, the solid phase of Fe at 33.385 g/cm$^3$ may be hexagonal
close-packed (HCP). However, the determinations of the lattice constants of HCP at high temperatures are very difficult, and the dynamical properties such as the structures and melting behavior are much similar to the BCC structures \cite{fe2,melt1}. Therefore, we used BCC structures as the initial configurations. In fact, from the tests of Fe at 45 g/cm$^3$, it is known that the transport properties are independent on the initial configurations when Fe melts. Furthermore, the effects of larger system sizes on calculated viscosities have been previously studied for models such as hard-sphere and Lennard-Jones liquids \cite{lj}, showing meaningful values for $\mu$ even with only 32 atoms. Thus, the initial configurations and sizes in this work can promise the reasonable accuracy of our simulations.

\section{Structural dynamics}

\begin{figure}[!tb]
\centering
\includegraphics*[width=3.2in]{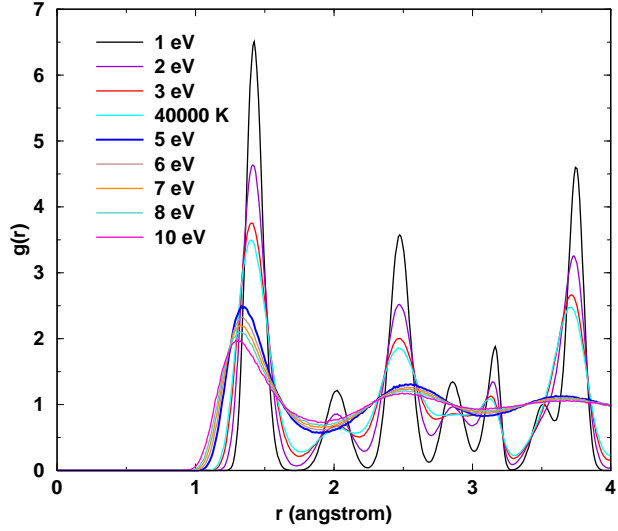}
\caption{(Color online) RDF vs temperature for Fe at 45 g/cm$^3$.} \label{fig3}
\end{figure}

\begin{figure}[!tb]
\centering
\includegraphics*[width=3.2in]{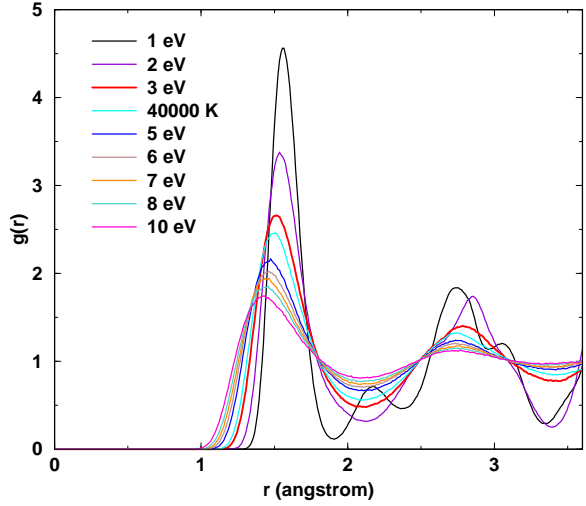}
\caption{(Color online) RDF vs temperature for Fe at 33.385 g/cm$^3$.} \label{fig4}
\end{figure}

The dynamical structures for dense matter are basic to understand the melting behaviors and the dynamics in planets core. However, dense liquid Fe is rarely studied before due to the lack of effective methods. Dense Fe can hold the solid phase or ordered structures at very high temperatures, form clusters and chemical bonds assisted by core electrons \cite{dai1}, and keep high melting temperature \cite{melt1,melt2,melt3}. Here, the structures are investigated by looking at the radial distribution functions (RDF) $g(r)$, as shown in Fig.~\ref{fig3} and Fig.~\ref{fig4}.

From the RDF, we can know that when the temperature increases up to 5 eV, Fe exhibits liquid behaviors at 45 g/cm$^3$, and above 3 eV at 33.385 g/cm$^3$. We calculated the details of the dynamical structures near 5 eV and 3 eV for the two densities respectively, using the two phase approach (TPA) \cite{melt1} with 216 atoms and 108 atoms to obtain the melting temperature. The melting temperatures are found to around 57000 K and 30000 K, within the errors of 200 K in our simulations. The same behaviors can be also found in the data of diffusion coefficients and viscosities, which are shown in Fig.~\ref{fig1} and Fig.~\ref{fig2} and will be discussed detailedly later. In this stage, the Fe at 45 g/cm$^3$ and 40000 K is also in solid phase in the Fe phase diagram \cite{fe2}.

We should note that the first peaks of the RDF move toward to the zero point, indicating the effective distance between ions decrease with the increasing temperatures. Therefore, we may use the positions of the first peaks in RDF as the effective atomic diameters when discussing about the SE relation later. If we use only the averaged ionic radius, the temperature effect can not be included since the radius is only dependent on the type of elements and system densities.

\section{Transport properties}

After the convergence tests, we can obtain the transport properties including EOS, diffusion coefficients and viscosities for dense Fe. The summaries of the data are shown in Table.~\ref{45g} and Table.~\ref{33g}. In the tables, pressures and diffusion coefficients from AAMD are also shown. When the solid phases are not broken, the diffusion coefficients are very small and the viscosities are very large (therefore they are not convergent in Green-Kubo equation).

\begin{figure}[!tb]
\centering
\includegraphics*[width=3.2in]{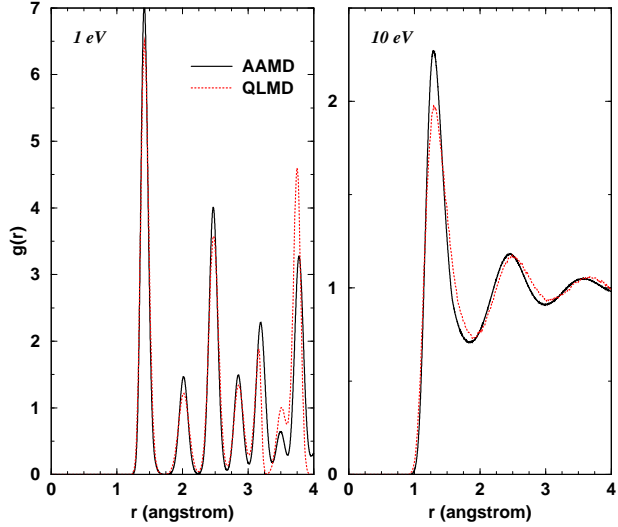}
\caption{(Color online) RDF comparison between AAMD and QLMD for Fe at 45 g/cm$^3$.} \label{fig5}
\end{figure}

For the density of 45 g/cm$^3$, from Table.~\ref{45g}, the diffusion coefficients change rapidly between the temperatures of 40000 K (3.447 eV) and 5 eV, and reach the orders of liquids diffusion at 5 eV. Also, the viscosities above 5 eV go to convergence when we integrate the pressure autocorrelations in Eq.~\ref{e3}, indicating the existence of liquid phases. However, AAMD gives the liquid Fe at 4 eV (not shown here), showing its limitations from pair potentials in the very strongly coupling regimes. Furthermore, we can see that the diffusion coefficients from Eq.~\ref{e1} and Eq.~\ref{e2} are equal, which shows the validation and convergence of our simulations. For the pressure, AAMD gives much larger pressures than those of QLMD simulations, which is caused by the electronic structure calculations from Thomas-Fermi methods. It is very interesting that when the temperatures are above 5 eV, the diffusion coefficients from QLMD and AAMD are very close. One reason might be that the diffusion coefficients are strongly dependent on the nearest neighbors distributions. Pair potentials can deal with the nearest neighbors interactions between ions with good accuracy. This can be verified using the comparisons of RDF from QLMD and AAMD, as shown in Fig.~\ref{fig5}, where the positions of the first peaks of RDF are almost overlapped. Therefore, the nearest neighbors interactions should be reasonable in AAMD. However, AAMD can not deal with many-body interactions, inducing the shear viscosities from AAMD are incorrect here, since dense Fe liquid holds a lot of medium ordered structures, and therefore the melting behavior is not correct within AAMD framework.

For the density of 33.385 g/cm$^3$, the behaviors are almost the same. The melting temperature locates below 3 eV, according to the RDF and diffusion coefficients. It is noticed that the pressures from AAMD are much closer to those from QLMD compared with the results of 45 g/cm$^3$. This means that the electronic structures from AA model are reasonable here but give more free electrons at 45 g/cm$^3$. Furthermore, the melting temperature from AAMD is also lower than that from QLMD, and the diffusion coefficients are in the same order when Fe melts above 3 eV.

\begin{center}
\begin{table}[!htb]
\caption{Summary of the results of dense Fe at different temperatures at 45 g/cm$^3$.
D$^R$ and D$^V$ are respectively defined in Eq.~\ref{e1} and \ref{e2}. D$_{AAMD}$ and
$\mu_{AAMD}$ are the diffusion coefficient and viscosity from AAMD calculations, respectively. \label{45g}} \centering
\begin{tabular}{c c c c c c c c}
\hline\hline
 T   &  P  & D$^R$   &  D$^V$   & $\mu$ & P$_{AAMD}$ & D$_{AAMD}$ &  $\mu_{AAMD}$ \\
(eV) &(Mbar)&($cm^2/s$ & ($cm^2/s$) & ($mPa\cdot s$) & (Mbar) & ($cm^2/s$) &  ($mPa\cdot s$) \\ \hline
 1      & 134.128  & 0.61$\times 10^{-7}$ & 0.60$\times 10^{-7}$  &  ---     & 245.13 & 2.39$\times 10^{-8}$  &  --- \\
 2      & 137.148  & 1.40$\times 10^{-7}$ & 1.61$\times 10^{-7}$  &  ---     & 250.25 & 9.55$\times 10^{-8}$  &  ---  \\
 3      & 140.594  & 2.24$\times 10^{-7}$ & 2.68$\times 10^{-7}$  &  ---     & 255.73   & 2.07$\times 10^{-7}$ & --- \\
 3.447  & 142.295  & 3.72$\times 10^{-7}$ & 3.82$\times 10^{-7}$  &  ---     & 258.11 & 2.91$\times 10^{-7}$  &  --- \\
 5      & 152.218  & 1.21$\times 10^{-4}$ & 1.23$\times 10^{-4}$  & 62.07$\pm$6  & 270.77 & 1.54$\times 10^{-4}$ &  104.29 \\
 6      & 156.825  & 2.22$\times 10^{-4}$ & 2.11$\times 10^{-4}$  & 44.11$\pm$5 & 275.11 & 2.11$\times 10^{-4}$ &  83.38 \\
 7      & 161.296  & 3.15$\times 10^{-4}$ & 3.05$\times 10^{-4}$  & 41.29$\pm$4 & 279.43  &  2.73$\times 10^{-4}$ & 69.47 \\
 8      & 166.081  & 3.26$\times 10^{-4}$ & 3.37$\times 10^{-4}$  & 38.13$\pm$4 & 283.65 & 3.39$\times 10^{-4}$  &  60.34 \\
 10     & 175.421  & 4.86$\times 10^{-4}$ & 4.83$\times 10^{-4}$  & 27.72$\pm$3  & 292.46 & 4.77$\times 10^{-4}$ &   48.64 \\
 \hline\hline
\end{tabular}
\end{table}
\end{center}

\begin{center}
\begin{table}[!htb]
\caption{Summary of the results of dense Fe at different temperatures at 33.385 g/cm$^3$. \label{33g}} \centering
\begin{tabular}{c c c c c c c c}
\hline\hline
 T   &  P  & D$^R$   &  D$^V$   & $\mu$ & P$_{AAMD}$ & D$_{AAMD}$ &  $\mu_{AAMD}$ \\
(eV) &(Mbar)&($cm^2/s$ & ($cm^2/s$) & ($mPa\cdot s$) & (Mbar) & ($cm^2/s$) &  ($mPa\cdot s$) \\ \hline
 1      & 60.547  & 3.50$\times 10^{-7}$ & 3.21$\times 10^{-7}$  &  ---     & 79.52 & 2.11$\times 10^{-6}$ &  --- \\
 2      & 63.471  & 2.45$\times 10^{-6}$ & 2.19$\times 10^{-6}$  &  ---     & 84.05 & 2.69$\times 10^{-5}$ &  --- \\
 3      & 67.739  & 6.70$\times 10^{-5}$ & 6.71$\times 10^{-5}$  & 47.1996$\pm$5 &  86.99  & 1.49$\times 10^{-4}$ & 29.43  \\
 3.447  & 69.786  & 1.16$\times 10^{-4}$ & 1.14$\times 10^{-4}$  & 40.3466$\pm$5 & 88.36 & 1.84$\times 10^{-4}$  & 27.32 \\
 5      & 75.282  & 2.98$\times 10^{-4}$ & 2.85$\times 10^{-4}$  & 24.217$\pm$3  & 92.57 & 3.28$\times 10^{-4}$  & 22.39 \\
 6      & 79.727  & 3.67$\times 10^{-4}$ & 3.74$\times 10^{-4}$  & 19.8484$\pm$2  & 95.33 & 4.15$\times 10^{-4}$  & 20.41\\
 7      & 82.248  & 5.99$\times 10^{-4}$ & 5.91$\times 10^{-4}$  & 18.1716$\pm$2  & 98.30 &  5.08$\times 10^{-4}$  & 19.00 \\
 8      & 85.825  & 6.08$\times 10^{-4}$ & 5.98$\times 10^{-4}$  & 17.1372$\pm$2  & 101.11 & 6.13$\times 10^{-4}$  & 17.83 \\
 10     & 93.098  & 7.51$\times 10^{-4}$ & 7.69$\times 10^{-4}$  & 15.7396$\pm$2  & 107.30 & 8.01$\times 10^{-4}$ &  16.28 \\
 \hline\hline
\end{tabular}
\end{table}
\end{center}

The small sizes in the calculations can introduce statistical errors. Especially for the viscosities, the errors are within 10\%. Therefore, calculations with more than 10000 atoms at least and accurate many-body potentials are needed basically. We need to construct more accurate semiclassical potentials for obtaining more accurate diffusion coefficients and viscosities.

\section{SE relation}

With respect to the simulation time and system sizes, we find that the diffusion coefficients can reach convergence much easier than the viscosities. Therefore, if we can obtain the viscosities from the calculated diffusions, a lot of computational resources can be saved. Therefore, the validation of SE relation is important and meaningful. In order to test the validation, we calculate viscosities from different definitions of effective atomic diameters: averaged ionic distance at a specific density $r_i=V^{(1/3)}$ (where V is the averaged volume of one atom at a specific density), the position of the first peaks in RDF ($r_f$), and the effective radius ($r_E$) from effective coordination number (ECN) definition. ECN can reveal the topology of the structures partly, and also give the local information around one atom. The definition of ECN is as following

\begin{equation} \label{ecn}
ECN=\frac{1}{N}\sum_{i=1}^NECN_i=\frac{1}{N}\sum_{i=1}^N\sum_{j\neq i}exp[1-(\frac{d_{ij}}{d_{av}^i})^6]
\end{equation}
where $d_{ij}$ is the distance between atoms $i$ and $j$; N is the total number of atoms in the system; $d_{av}^i$ and its average value $d_{av}$ are defined as
\begin{equation} \label{dav}
d_{av}^i=\frac{\sum_jd_{ij}exp[1-(\frac{d_{ij}}{d_{av}^i})^6]}{\sum_jexp[1-(\frac{d_{ij}}{d_{av}^i})^6]}, d_{av}=\frac{1}{N}\sum_{i=1}^Nd_{av}^i
\end{equation}

With this definition, the effective diameters $r_E = d_{av}$, which includes the temperature and density effects.

The results of different effective atomic diameters and the calculated viscosities from SE relation are shown in Table.~\ref{se1} and Table.~\ref{se2}. Comparing the results with the calculated viscosities from Green-Kubo equation, we find that the viscosities are very different at low temperatures. With the increasing temperature, SE relation seems to be more and more validated. Moreover, the choice of the effective atomic diameters plays an important role in the validation of the SE relation, where definitions of the positions of the first peaks of RDF and diameters from ECN calculations are more appropriated. Considering the calculation uncertainties and statistical errors, the SE relations can be considered validated only when the choice of the effective diameters are reasonable.

\begin{center}
\begin{table}[!tb]
\caption{Summary of the effective diameters and its corresponding viscosities for dense Fe at different temperatures at 45 g/cm$^3$. $\overline{Z}$ is the average ionization degree from AAMD calculations. $\mu_i$, $\mu_f$, $\mu_E$ are the viscosities corresponding to different definitions of effective atomic diameters of ions $r_i$, $r_f$ and $r_E$, respectively.  $\mu_{YVM}$ is the viscosity from the model of Yukawa viscosity model (YVM) (see details in Ref.~\cite{murrilo}.) \label{se1}}
\centering
\begin{tabular}{c c c c c c c c c}
\hline\hline
 T   &  $r_i$  & $\mu_i$  &  $r_f$  & $\mu_f$ & $r_E$ & $\mu_E$ & $\overline{Z}$ & $\mu_{YVM}$ \\
(eV) & ({\AA}) & ($mPa\cdot s$) & ({\AA}) & ($mPa\cdot s$) & ({\AA}) &($mPa\cdot s$) & & ($mPa\cdot s$)  \\ \hline
   5    & 1.275      &          83.361      &   1.335         &         79.612       & 1.338 & 79.428  & 13.011 & 43.99 \\
   6    & 1.275      &          54.002      &   1.330         &         51.767       & 1.323 & 52.052  & 13.009 & 37.97 \\
   7    & 1.275      &          46.527      &   1.325         &         44.770       & 1.309 & 45.302  & 13.009 & 33.77 \\
   8    & 1.275      &          49.051      &   1.320         &         47.377       & 1.297 & 48.219  & 13.002 & 30.71 \\
   10   & 1.275     &           41.166      &   1.295         &         40.529       & 1.276& 41.127   & 12.993 & 26.64 \\
 \hline\hline
\end{tabular}
\end{table}
\end{center}

\begin{center}
\begin{table}[!tb]
\caption{Summary of the effective diameters and its corresponding viscosities for dense Fe at different temperatures at 33.385 g/cm$^3$.\label{se2}} \centering
\begin{tabular}{c  c  c c c  c  c  c c}
\hline\hline
 T   &  $r_i$  & $\mu_i$  &  $r_f$  & $\mu_f$ & $r_E$ & $\mu_E$ & $\overline{Z}$ & $\mu_{YVM}$ \\
(eV) & ({\AA}) & ($mPa\cdot s$) & ({\AA}) & ($mPa\cdot s$) & ({\AA}) &($mPa\cdot s$) & & ($mPa\cdot s$)  \\ \hline
   3    & 1.408      &          81.002      &   1.515         &         75.301       & 1.488& 76.667  & 9.344 & 28.65 \\
   3.447 & 1.408      &         53.831      &   1.505         &         50.375       & 1.473 & 51.459 & 9.343 & 25.81 \\
   5    & 1.408      &          30.300      &   1.475         &         28.931       & 1.437 & 29.687 & 9.332 & 20.23 \\
   6    & 1.408      &          29.633      &   1.465         &         28.488       & 1.419 & 29.410 & 9.326 & 18.41 \\
   7    & 1.408      &          21.147      &   1.455         &         20.469       & 1.404 & 21.214 & 9.323 & 17.28 \\
   8    & 1.408      &          23.847      &   1.450         &         23.162       & 1.388 & 24.197 & 9.312 & 16.56 \\
   10   & 1.408     &           24.122      &   1.425         &         23.840       & 1.364& 24.912  & 9.296 & 15.91 \\
 \hline\hline
\end{tabular}
\end{table}
\end{center}

In order to avoid the usage of SE relation, some models have been constructed. For example, Murrilo reported the construction of the Yukawa viscosity model (YVM) based on molecular dynamics simulations \cite{murrilo}, which can be used in warm dense regime. We verify the validation of YVM model here. First of all, we give the average ionization degree ($\overline{Z}$) from AA model with Electronic Energy-level Broadening \cite{aamd1}, and then show the viscosities ($\mu_{YVM}$) from YVM model, as shown in Table.~\ref{se1} and Table~\ref{se2}. It can be shown that YVM model improve the accuracy of SE relation, especially for the cases at high temperatures. This suggests that the Yukawa potential might be reasonable for the states at so high density at relatively high temperatures.

\begin{figure}[!tb]
\centering
\includegraphics*[width=3.2in]{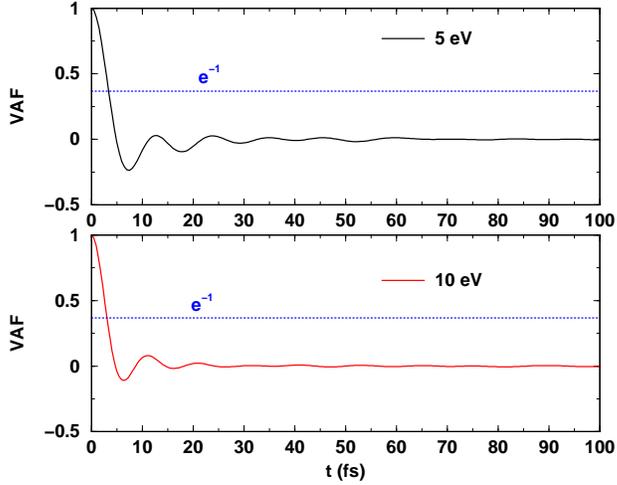}
\caption{(Color online) Velocity autocorrelation function (VAF) of Fe ions at the temperatures of 5 eV (upper panel)) and 10 eV (lower panel) from QLMD calculations for Fe at 45 g/cm$^3$.} \label{fig6}
\end{figure}

What's the possible reason for the debatable validation of SE relation in this regime? The basic physics is whether the ions exhibit as Brownian particles. For the specific case in this work, heavy Fe ions are moving in the liquid-like electron sea, randomly collided with many free-like electrons. However, the other dominant factor for the ionic motions are the interactions between ions, i.e., the strong coupling of ions. Every ion moves in a specific force fields determined by both electrons and ions, and this force field is not constant. The ions in dense matter have memory effect, i.e., the ionic positions at this time plays important role in the ionic movements at next time. In other words, if the correlation time of ions is larger than the observable time, the motion of the ions should not be Morkov process. In order to analyze the correlation time of the systems, the velocity autocorrelation functions (VAF) of ions at 5 eV and 10 eV are shown in Fig.~\ref{fig6}. Here, we adopt the simplest definition of the correlation time ($\tau_c$) to compare the correlation time at different temperatures. Here, $\tau_c$ is defined as the time for which the VAF(t) decreases to $1/e$ \cite{cor}. With this definition, $\tau_c$ is about 4.1 fs and 2.7 fs for the temperatures of 5 eV and 10 eV, respectively. Therefore, with the increasing temperature, the behaviors of ions are more and more Brownian-like motion. In fact, if the system is ideal gas, the correlation time will be zero, indicating uncorrelated behaviors between two near time steps. That is to say, when temperature is high enough, the random collisions between particles are dominant, and the statistical behaviors can work well. If the correlation time of the system is zero, i.e, we can recover SE relation directly from Eq.~\ref{e2} and Eq.~\ref{e3} \cite{bookMD}. We can then conclude that the SE relation is strongly dependent on the coupling parameters of ions $\Gamma=Z^{*2}/(k_BTa)$, with T the system temperature, $k_B$ the Boltzmann constant, $a$ the mean ionic sphere radius defined as $a=(3/(4{\pi}n_i))^{1/3}$, $Z^*$ the average ionization degree, $n_i$ the ionic number density. If $\Gamma$ is small enough, SE relation can be valid. Furthermore, one possible reason that QMD or QLMD can obtain the reasonable transport properties within relatively short time (within 10 ps) is the short correlation time of particles as shown in Fig.~\ref{fig6}.

\section{Conclusion}

In conclusion, the dynamical structures and transport properties including EOS, diffusion and viscosities are calculated using QLMD method within the framework of first principles. The dynamical structures show the information of RDF, melting behaviors, diffusions and viscosities. The validation of SE relation is also discussed, showing the importance of the methods of choosing the effective atomic diameters. This work studies the dense liquid Fe existed in giant planets such as Jupiter, exoplanets, indicating the necessity of first principles calculations or constructing accurate many-body interactions. Furthermore, the results are crucial for understanding the laser-shock induced compression experiments, reminding us of the complexity of dense matters. In this field, a lot of new physics needs to be deeper studied.

\section*{Acknowledgments}
This work is supported by the National NSFC under Grant Nos. 11104351, 11274383, 60921062, 11005153 and 11104350. Calculations are carried out at the Research Center of Supercomputing Application, NUDT.


\section*{References}
\begin{thebibliography}{9}
\bibitem{apj1} D. C. Swift \textit{et al} 2012 Mass-radius realtionships for exoplanets. \textit{Astrophys. J.} {\bf 744} 59.
\bibitem{apj2} Jin F T \textit{et al} Radiative opacity of iron studtied using a detailed level acounting model. \textit{Astrophys. J.} {\bf 693} 597¨C609.
\bibitem{gao} Gao C and Zeng J L. 2008 Spectrally resolved and Rosseland and Planck mean opacities of iron plasmas at temperatures above 100 eV: A systematic study. \textit{Phys. Rev. E} {\bf 78} 046407.
\bibitem{nif} Moses E I \textit{et al.} 2009 The National Ignition Facility: Ushering in a new age for high energy density science. \textit{Phys. Plasmas} {\bf 16} 041006.
\bibitem{laser} R. Jeanloz \textit{et al.} 2007 Achieving high-density states through shock-wave loading of precompressed samples. \textit{Proc. Natl. Acad. Sci. U.S.A.} {\bf 104} 9172-9177.
\bibitem{fe1} Pickard C J and Needs R J 2009 Stable phases of iron at terapascal pressures \textit{J. Phys.: Condens. Matter} {\bf 21} 452205.
\bibitem{fe2} Stixrude L 2012 Structure of Iron to 1 Gbar and 40 000 K. \textit{Phys. Rev. Lett.} {\bf 108} 055505.
\bibitem{tran1} Wijs Gilles A \textit{et al} 1998 The viscosity of liquid iron at the physical conditions of the Earth¡¯s core \textit{Nature} {\bf 392} 805.
\bibitem{tran2} Alf\`{e} D, Gillan M J and Price D G 1999 The melting curve of iron at the pressures of the Earth's core from ab initio calculations. \textit{Nature} {\bf 401} 462.
\bibitem{tran3} Alf\`{e} D, and Gillan M J 1998 First-Principles Calculation of Transport Coefficients. \textit{Phys. Rev. Lett.} {\bf 81} 5161.
\bibitem{tran4} Alf\`{e} D, and Gillan M J 1998 First-principles simulations of liquid Fe-S under Earth¡¯s core conditions. \textit{Phys. Rev. B} {\bf 58} 8248.
\bibitem{tran5} Vo\v{c}adlo L, Alf\`{e} D, Price G D and Gillan M J 2000 Transport coeffcients of liquids from first principles. \textit{Physics of the Earth and Planetary Interiors} {\bf 120} 145.
\bibitem{Saigo} Saigo T, and Hamaguchi S 2002 Shear viscosity of strongly coupled Yukawa systems. \textit{Phys. Plasmas} {\bf 9} 1210-1216.
\bibitem{murrilo} Murillo M S 2008 Viscosity estimates of liquid metals and warm dense matter using the Yukawa reference system. \textit{High Energy Density Phys.} {\bf 4} 49-57.
\bibitem{Daligault} Daligault J 2012 Diffusion in Ionic Mixtures across Coupling Regimes. \textit{Phys. Rev. Lett} {\bf 108} 225004.
\bibitem{dense1} Ma Y M \textit{et al}. 2009 Transparent dense sodium. \textit{Nature} {\bf458} 182.
\bibitem{dense2} Pickard C J, and Needs R J 2010 Aluminium at terapascal pressures. Nature Mater. {\bf 9} 624-6227.
\bibitem{dai1} Dai J Y \textit{et al} 2011 Dynamical Ionic Clusters with Flowing Electron Bubbles from Warm to Hot Dense Iron along the Hugoniot Curve. \textit{Phys. Rev. Lett.} {\bf 109} 175701.
\bibitem{dai2} Dai J Y, Hou Y, and Yuan J M 2010 Unified first principles description from warm dense matter to ideal ionized gas plasma: Electron-ion induced friction. \textit{Phys. Rev. Lett.} {\bf 104} 245001.
\bibitem{dai4} Dai J Y, Hou Y, and Yuan J M 2010 Quantum Langevin molecular dynamics determination of the solar-interior equation of state. \textit{Astrophys. J.} {\bf 721} 1158-1163.
\bibitem{aayuan} Yuan J 2002 Self-consistent average-atom scheme for electronic structure of hot and dense plasmas of mixture. \textit{Phys. Rev. E} {\bf 66} 047401.
\bibitem{aamd1} Hou Y, Jin F T and Yuan J M 2006 Influence of the electronic energy level broadening on the ionization of atoms in hot and dense plasmas: An average atom model demonstration. \textit{Phys. Plasmas} {\bf 13} 093301 and references therein.
\bibitem{TF1} Mazevet S, Lambert F, Bottin F, Z\'{e}rah G, and Cl\'{e}rouin J 2007 Ab initio molecular dynamics simulations of dense boron plasmas up to the semiclassical Thomas-Fermi regime. \textit{Phys. Rev. E} {\bf 75} 056404.
\bibitem{TF} Lambert F, Cl\'{e}rouin J, and Z\'{e}rah G 2006 Very-high-temperature molecular dynamics \textit{Phys. Rev. E} {\bf 73} 016403.
\bibitem{TF2} Danel J-F, Kazandjian L, and Z\'{e}rah G 2012 Numerical convergence of the self-diffusion coefficient and viscosity obtained with Thomas-Fermi-Dirac molecular dynamics \textit{Phys. Rev. E} {\bf 85} 066701.
\bibitem{Mermin} Mermin N D 1965 Thermal Properties of the Inhomogeneous Electron Gas. \textit{Phys. Rev.} {\bf 137} A1441-A1443.
\bibitem{s2} Graziani F R \textit{et al} 2012 Large-scale molecular dynamics simulations of dense plasmas: The Cimarron project. \textit{High Energy Density Phys.} {\bf 8} 105-131.
\bibitem{dai5} Dai J Y, Hou Y, and Yuan J M 2011 Influence of ordered structures on electrical conductivity and XANES from warm to hot dense matter. \textit{High Energy Density Phys.} {\bf 7} 84-87.
\bibitem{wdm2} Kress J D, Cohen James S, Horner D A, Lambert F, and Collins L A 2010 Viscosity and mutual diffusion of deuterium-tritium mixtures in the warm-dense-matter regime. \textit{Phys. Rev. E} {\bf 82} 036404.
\bibitem{LiH} Horner D A, Lambert F, Kress J D, and Collins L A 2009 Transport properties of lithium hydride from quantum molecular dynamics and orbital-free molecular dynamics \textit{Phys. Rev. B} {\bf 80} 024305.
\bibitem{CH} Lambert F, and Recoules V 2012 Plastic ablator and hydrodynamic instabilities: A first-principles set of microscopic coefficients \textit{Phys. Rev. E} {\bf 86} 026405.
\bibitem{Pu} Kress J D \textit{et al} 2011 Quantum molecular dynamics simulations of transport properties in liquid and dense-plasma plutonium. \textit{Phys. Rev. E} {\bf 83} 026404.
\bibitem{dai3} Dai J Y and Yuan J M 2009 Large scale efficient Langevin dynamics, and why it works. \textit{Europhys. Lett.} {\bf 88} 20001.
\bibitem{qe} Giannozzi P \textit{et al} 2009 QUANTUM ESPRESSO: a modular and open-source software project for quantum simulations of materials. \textit{J. Phys.: Cond. Matter} {\bf 21} 395502.
\bibitem{gga} Perdew J P, Burke K, and Ernzerhof M 1996 Generalized gradient approximation made simple. \textit{Phys. Rev. Lett.} {\bf 77} 3865-3868.
\bibitem{bulk} Zhang H L \textit{et al} 2010 Static equation of state of bcc iron. \textit{Phys. Rev. B} {\bf 82} 132409.
\bibitem{aamd2} Hou Y, and Yuan J M 2009 Alternative ion-ion pair-potential model applied to molecular dynamics simulations of hot and dense plasmas: Al and Fe as examples. \textit{Phys. Rev. E} {\bf 79} 016402.
\bibitem{gk} Allen M P and Tildesley T J 1987 Computer Simulation of
Liquids. \textit{Clarendon, Oxford}.
\bibitem{wdm1} Collins L, Kwon I, Kress J, Troullier N, and Lynch D 1995 Quantum molecular dynamics simulations of hot, dense hydrogen. \textit{Phys. Rev. E} {\bf 52} 6202-6219.
\bibitem{wdm3} Mithen J P, Daligault J, and Gregori G 2012 Molecular Dynamics Simulations for the Shear Viscosity of the One-Component Plasma. \textit{Contrib. Plasma Phys.} {\bf 52} 58-61.
\bibitem{wdm5} Lorenzen W, Holst B, and Redmer R 2009 Demixing of Hydrogen and Helium at Megabar Pressures. \textit{Phys. Rev. Lett.} {\bf 102} 115701.
\bibitem{wdm6} Plagemann K U \textit{et al} 2012 Dynamic structure factor in warm dense beryllium. \textit{New J. Phys} {\bf 14} 055020.
\bibitem{SE1} Cappelezzo M, Capellari C A, Pezzin S H, and Coelho L A F 2007 Stokes-Einstein relation for pure simple fluids. \textit{J. Chem. Phys.} {\bf 126} 224516.
\bibitem{lj} Schoen M and Hoheisel C 1985 The shear viscosity of a Lennard-Jones fluid calculated by equilibrium molecular dynamics. \textit{Mol. Phys.} {\bf 56} 653-672.
\bibitem{melt1} Morard G, Bouchet J, Valencia D, Mazevet S, Guyot F 2011 The melting curve of iron at extreme pressures: Implications for planetary cores. \textit{High Energy Density Phys.} {\bf 7} 141-144.
\bibitem{melt2} Nguyen Jeffrey H, and Holmes Neil C 2004 Melting of iron at the physical conditions of the Earth¡¯s core. \textit{Nature} {\bf 427} 339-342.
\bibitem{melt3} Alf\`{e} D 2009 Temperature of the inner-core boundary of the Earth: Melting of iron at high pressure from first-principles coexistence simulations. \textit{Phys. Rev. B} {\bf 79} 060101.
\bibitem{cor} Kesselring T A, Franzese G, Buldyrev S V, Herrmann H J and Stanley H E 2012 Nanoscale Dynamics of Phase Flipping in Water near its Hypothesized Liquid-Liquid Critical Point. \textit{Sci. Rep.} {\bf 2} 474.
\bibitem{bookMD} Frenkel D and Smit B 1996 Understanding molecular simulation. \textit{Acedemic Press, San Diego, USA}.
\end{thebibliography}
\end{document}